\documentclass{aastex62}

\usepackage{graphicx}
\usepackage{graphics}
\usepackage{amssymb}
\usepackage{bm}
\usepackage{times}

\shorttitle{Massive DLAs at Intermediate Redshifts}
\shortauthors{Kanekar et al.}
\begin{document}
\title{Massive, Absorption-selected Galaxies at Intermediate Redshifts}

\correspondingauthor{Nissim Kanekar}
\email{nkanekar@ncra.tifr.res.in}

\author{N. Kanekar}, 
\altaffiliation{DST Swarnajayanti Fellow}
\affiliation{National Centre for Radio Astrophysics, Tata Institute of Fundamental Research, 
Pune University, Pune 411007, India}

\author{J. X. Prochaska}
\affiliation{University of California Observatories-Lick Observatory, University of California, Santa Cruz, CA, 95064, USA}

\author{L.~Christensen}
\affiliation{Dark Cosmology Centre, Niels Bohr Institute, Copenhagen University, Juliane Maries Vej 30, 2100 Copenhagen O, Denmark}

\author{N.~H.~P. Rhodin}
\affiliation{Dark Cosmology Centre, Niels Bohr Institute, Copenhagen University, Juliane Maries Vej 30, 2100 Copenhagen O, Denmark}

\author{M. Neeleman}
\affiliation{University of California Observatories-Lick Observatory, University of California, Santa Cruz, CA, 95064, USA}
\affiliation{Max-Planck-Institut f{\"u}r Astronomie, K{\"o}nigstuhl 17, D-69117 Heidelberg, Germany}

\author{M. A. Zwaan}
\affiliation{European Southern Observatory, Karl-Schwarzschildstrasse 2, 85748 Garching Bei Muenchen, Germany}

\author{P. M{\o}ller}
\affiliation{European Southern Observatory, Karl-Schwarzschildstrasse 2, 85748 Garching Bei Muenchen, Germany}

\author{M. Dessauges-Zavadsky}
\affiliation{Observatoire de Gen{\`e}ve, Universit{\'e} de Gen{\`e}ve, 51 Ch. des Maillettes, 1290 Sauverny, Switzerland}

\author{J.~P.~U. Fynbo}
\affiliation{The Cosmic Dawn Center, Niels Bohr Institute, Copenhagen University, DK-2100 Copenhagen, Denmark}

\author{T. Zafar}
\affiliation{6 Australian Astronomical Observatory, PO Box 915, North Ryde, NSW 1670, Australia}

\begin{abstract}
The nature of absorption-selected galaxies and their connection to the general galaxy population have 
been open issues for more than three decades, with little information available on their gas properties. 
Here we show, using detections of carbon monoxide (CO) emission with the 
Atacama Large Millimeter/submillimeter Array (ALMA), that five of seven high-metallicity, absorption-selected 
galaxies at intermediate redshifts, $z \approx 0.5-0.8$, have large molecular gas masses, 
$M_{\rm Mol} \approx (0.6 - 8.2) \times 10^{10} \: {\rm M}_\odot$ and high molecular gas fractions 
($f_{\rm Mol} \equiv \: M_{\rm Mol}/(M_\ast + M_{\rm Mol}) \approx 0.29-0.87)$. Their modest star formation 
rates (SFRs), $\approx (0.3-9.5) \: {\rm M}_\odot$~yr$^{-1}$, then imply long gas depletion timescales, 
$\approx (3 - 120)$~Gyr. The high-metallicity absorption-selected galaxies at $z \approx 0.5-0.8$ appear 
distinct from populations of star-forming galaxies at both $z \approx 1.3-2.5$, during the peak of star 
formation activity in the Universe, and lower redshifts, $z \lesssim 0.05$.  Their relatively low SFRs, 
despite the large molecular gas reservoirs, may indicate a transition in the nature of star formation 
at intermediate redshifts, $z \approx 0.7$.
\end{abstract}


\section{Introduction} 
\label{sec:intro}

At cosmological distances, galaxies are usually identified by their stellar emission, which causes 
them to appear bright at rest-frame ultraviolet (UV) and optical wavelengths. Such ``emission-selected'' 
samples have been used to glean much information on galaxy evolution, including the detection of galaxy 
populations at high redshifts, $z \approx 10$ \citep{oesch14}, the redshift evolution of the cosmic SFR 
density \citep{bouwens14} and the UV galaxy luminosity function \citep{bouwens15}, relations between 
the stellar mass of a galaxy and its SFR and metallicity \citep{tremonti04,noeske07}, etc. However, such 
samples, selected in the rest-frame UV or optical, contain an intrinsic bias towards brighter galaxies, 
and against dusty galaxies (where the dust obscures the optical or UV emission).

An alternative way of identifying high-redshift galaxies, free from this luminosity bias, is to select 
them based on their absorption signatures in the spectra of higher-redshift quasi-stellar objects (QSOs).
The highest H{\sc i} column density absorption systems ($N_{\rm HI} \geq 2 \times 10^{20}$~cm$^{-2}$), 
the damped Lyman-$\alpha$ absorbers \citep[DLAs;][]{wolfe05}, have long been identified with galaxies at 
high redshifts, because their H{\sc i} column densities are similar to those of galaxies in the 
local Universe and their metallicities are higher than those in the intergalactic medium, 
indicating enrichment via star formation \citep{wolfe05}. Recent searches through the Sloan Digital Sky 
Survey have revealed more than 10,000 DLAs at $z \gtrsim 2$ \citep{noterdaeme12b}.

Characterizing the galaxies that give rise to DLAs and connecting them to the emission-selected galaxy 
population is of great value in studies of galaxy evolution. Absorption studies of DLAs have 
yielded detailed information about the pencil beam traced by the QSO sightline, including the metallicity 
\citep{rafelski12}, kinematics \citep{prochaska97}, gas temperature \citep{kanekar14}, and star formation history 
\citep{dessauges07}. However, despite many 
searches, a combination of the bright background QSO and the intrinsic faintness of most high-$z$ DLA 
hosts has made it hard to detect the stellar emission from the galaxies associated with DLAs 
at high redshifts, $z \gtrsim 2$ \citep[e.g.][]{fynbo13,fumagalli15,krogager17}.
At intermediate redshifts, $ z \approx 0.5-1$, optical attempts to identify the DLA hosts have 
been more successful, and suggest that the absorbers probe gas in and around star-forming, field galaxies
\citep[e.g.][]{lebrun97,rao03,chen05}. 

Until recently, we have had no information on a vital part of the puzzle, the molecular gas in the 
DLA galaxies that gives rise to star formation. The advent of ALMA has changed the field, with 
the detections of CO emission from galaxies associated with a $z\approx 0.101$ sub-DLA \citep{neeleman16b}, 
a $z \approx 0.7163$ DLA \citep{moller18}, a $z \approx 0.633$ Lyman-limit absorber \citep{klitsch18}, 
and a $z \approx 2.193$ DLA \citep{neeleman18}, and of the $157.74\mu$m fine-structure 
transition of singly-ionized carbon from galaxies associated with two DLAs 
at $z \approx 4$ \citep{neeleman17}. \citet{neeleman16b} found the host of the 
$z = 0.101$ sub-DLA to be a molecule-rich, rotating disk, while \citet{moller18} found the 
galaxy associated with the $z = 0.7163$ DLA to be a massive, molecule-rich galaxy, but with a low SFR. 
We report here on an ALMA search for redshifted CO(2$-$1) emission from a sample of seven 
absorption-selected galaxies at $z \approx 0.5-0.8$, selected from known DLAs and sub-DLAs at these 
redshifts \citep{rao06} based on their high metallicity, $\approx 0.5-1.5$ times the solar metallicity. 
All our targets have galaxies identified at the absorption redshift and close to the quasar sightline, 
with impact parameter $\lesssim 50$~kpc \citep{chen05,christensen14,straka16,moller18}.\footnote{We 
use a flat $\Lambda$ cold dark matter cosmology throughout this paper, defined by the 
parameters, $\Omega_{\Lambda}=0.69$, $\Omega_{\rm M}=0.31$ and $H_0 = 71$~km~s$^{-1}$ \citep{planck16}.}

\section{Observations and Data analysis} 

\setcounter{table}{0}
\begin{table*}
\centering
\caption{The results.
\label{tab:table1}}
\begin{tabular}{|c|c|c|c|c|c|c|c|c|}
\hline
QSO & $z_{\rm QSO}$ & $z_{\rm abs}$$^a$ & N$_{\rm HI}$               & Beam$^b$ & RMS$_{\rm CO}$$^c$  & $\int S_{\rm CO} d{\rm V}$$^d$ & W20$^e$ & $L'_{\rm CO(1-0)}$$^f$ \\
	&           &                   & $\times 10^{20}$~cm$^{-2}$ & $'' \times ''$ & $\mu$Jy &     Jy~km/s   &  km/s  &  $\times 10^9$~K~km/s~pc$^2$ \\
\hline
\hline
B0827+243  & $0.941$ & $0.5247$ & $2.0$  & $2.1 \times 1.6$ & 275 & $3.11 \pm 0.16$   & 425 & $19.06 \pm 0.98$ \\
B1629+120  & $1.795$ & $0.5313$ & $5.0$  & $2.7 \times 2.4$ & 200 & $0.243 \pm 0.040$ & 350 & $1.53 \pm 0.25$  \\
J0058+0155 & $1.954$ & $0.6125$ & $1.1$  & $2.9 \times 2.0$ & 155 & $< 0.089$         & $-$ & $<0.75$          \\
J2335+1501 & $0.791$ & $0.6798$ & $0.5$  & $2.7 \times 2.5$ & 185 & $0.601 \pm 0.064$ & 350 & $6.26 \pm 0.67$  \\
J1323-0021 & $1.392$ & $0.7163$ & $3.5$  & $2.7 \times 2.2$ & 175 & $0.50 \pm 0.05$   & 650 & $5.79 \pm 0.58$  \\
J1436-0051 & $1.275$ & $0.7377$ & $1.2$  & $2.8 \times 2.2$ & 150 & $1.480 \pm 0.055$ & 350 & $18.19 \pm 0.68$ \\
J0138-0005 & $1.341$ & $0.7821$ & $0.65$ & $2.9 \times 2.1$ & 155 & $<0.082$          & $-$ & $< 1.1$          \\
\hline
\end{tabular}
\vskip 0.1in
$^a$~The absorber redshift, from low-ionization metal lines.\\
$^b$~The full-width-at-half-maximum of the ALMA synthesized beam.\\
$^c$~The root-mean-square (RMS) noise, on the final spectral cube, at a velocity resolution of $100$~km/s.\\
$^d$~The velocity-integrated CO(2$-$1) line flux density, or the $3\sigma$ upper limit on this quantity for the two CO non-detections, assuming a CO line FWHM of 300~km/s.\\
$^e$~The velocity width of the CO(2$-$1) line between points that are 20\% of the maximum value, 
for the CO detections (uncorrected for galaxy inclination).\\
$^f$~The inferred CO(1$-$0) line luminosity.\\
\vskip 0.1in
\end{table*}

The Atacama Millimeter/sub-millimeter Array (ALMA) was used to observe the seven target 
absorbers over 2014~December to 2017~January (programme ID's 2013.1.01178.S and 2015.1.00561.S), 
using the ALMA Band-4 receivers. All observations used four ALMA 1.875~GHz bands, with one band (sub-divided 
into 480~channels, using the FDM mode of the correlator) covering the expected redshifted CO(2$-$1) line 
frequency, and the other three bands (sub-divided into 128~channels, using the TDM mode) placed at 
neighbouring frequencies to measure the continuum emission. This yielded a velocity resolution of 
$\approx 8$~km/s for the band covering the redshifted CO(2$-$1) line, and a total velocity coverage of $\approx 4000$~km/s.

The initial calibration of all data was done by the ALMA support staff, using the 
ALMA data pipeline in the Common Astronomy Software Applications (CASA) package \citep{mcmullin07}. 
In two cases (B1629+120 and B0827+243), the quasar was sufficiently bright to allow us to 
self-calibrate the continuum data, which was carried out in the Astronomical Image Processing System 
\citep{greisen03}. The root-mean-square 
(RMS) noise on the continuum images was $\approx 13-20 \mu$Jy, except for the field 
of B0827+243, where the RMS noise was higher ($\approx 40 \mu$Jy), probably due to dynamic range limitations, 
as the quasar is very bright, with a flux density of $\approx 568$~mJy at the observing frequency. The 
antenna-based gains obtained from the procedure were then applied to the CO(2$-$1) line data, and the 
continuum image subtracted out to produce a residual visibility data set. This was then imaged in the CASA 
package to produce the final spectral cubes. The cubes were made at velocity resolutions of $50-300$~km/s, 
and used natural weighting, to maximize the signal-to-noise ratio. A correction for the ALMA primary 
beam was applied to the cubes, before carrying out the search for redshifted CO(2$-$1) emission. 

Redshifted CO(2$-$1) line emission was detected in five of our seven target fields (see Table~\ref{tab:table1}); the final CO 
spectra were obtained by measuring the flux density per channel in an elliptical region chosen 
to contain all detected emission. The $z= 0.7163$ DLA towards J1323-0021 was the only system detected in its 
rest-frame 230~GHz continuum emission, at $\approx 3.2\sigma$ significance with a flux
density of $53 \mu$Jy; this system is discussed in detail in \citet{moller18}. 

\begin{figure*}[t!]
\centering
\includegraphics[width=2.2in]{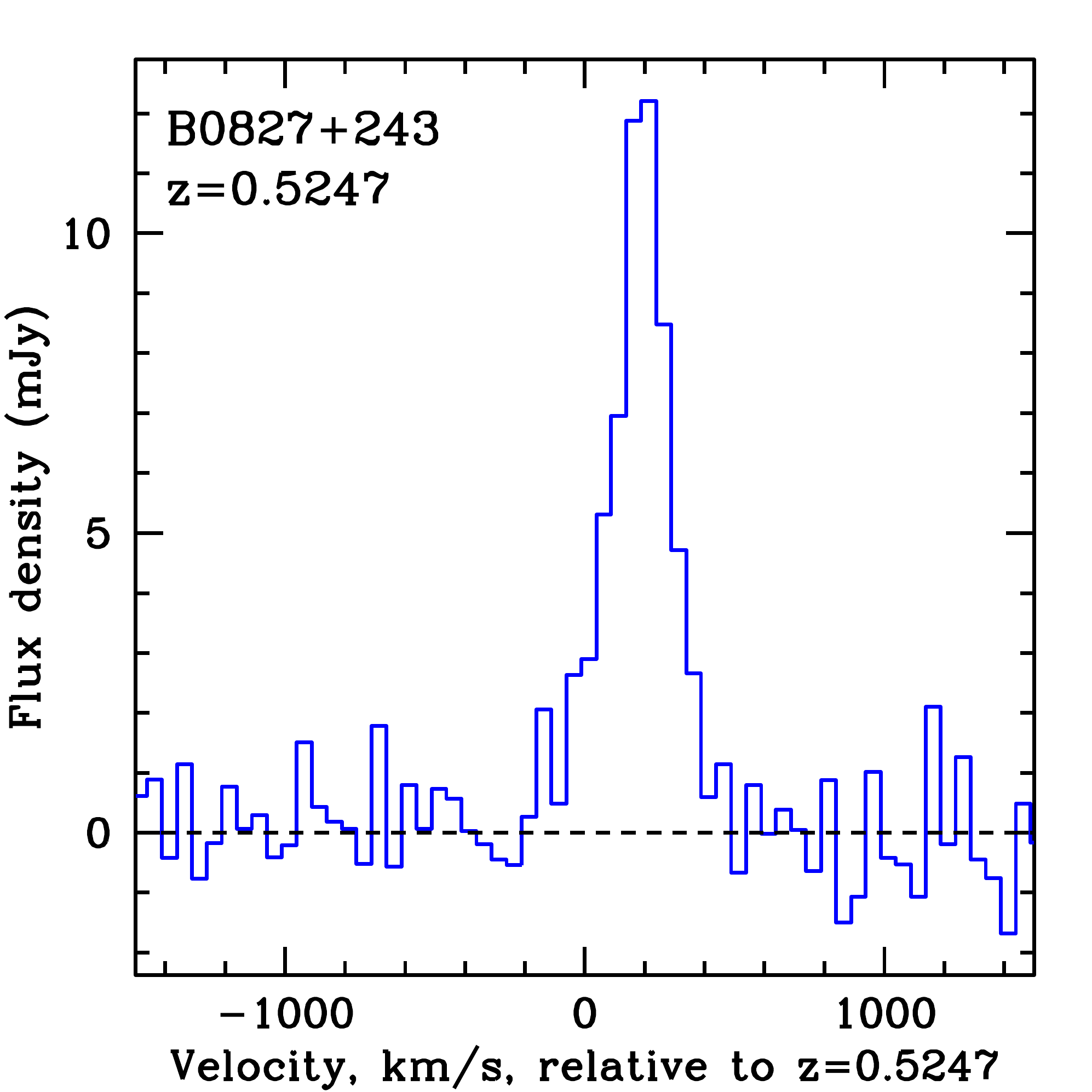}
\includegraphics[width=2.2in]{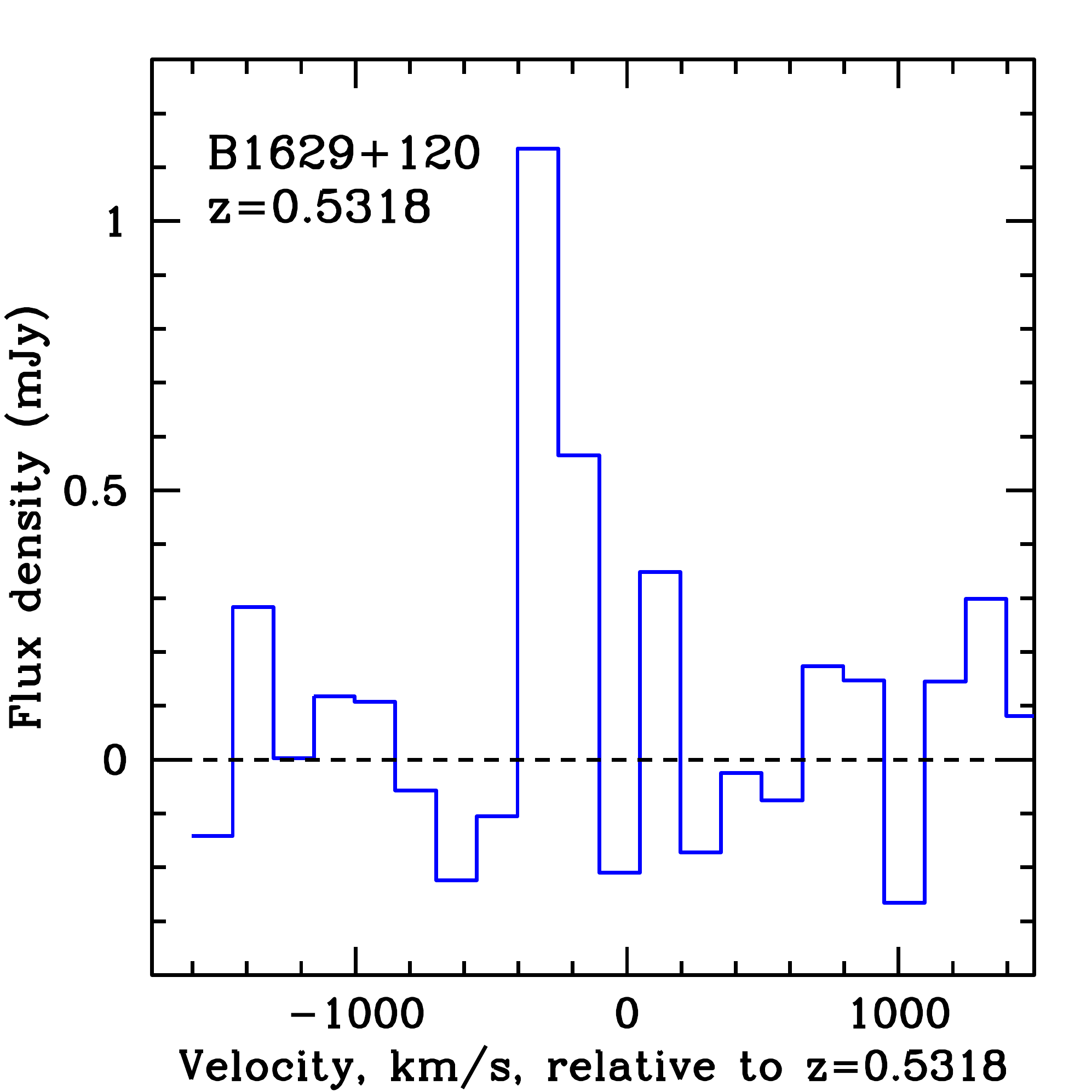}
\includegraphics[width=2.2in]{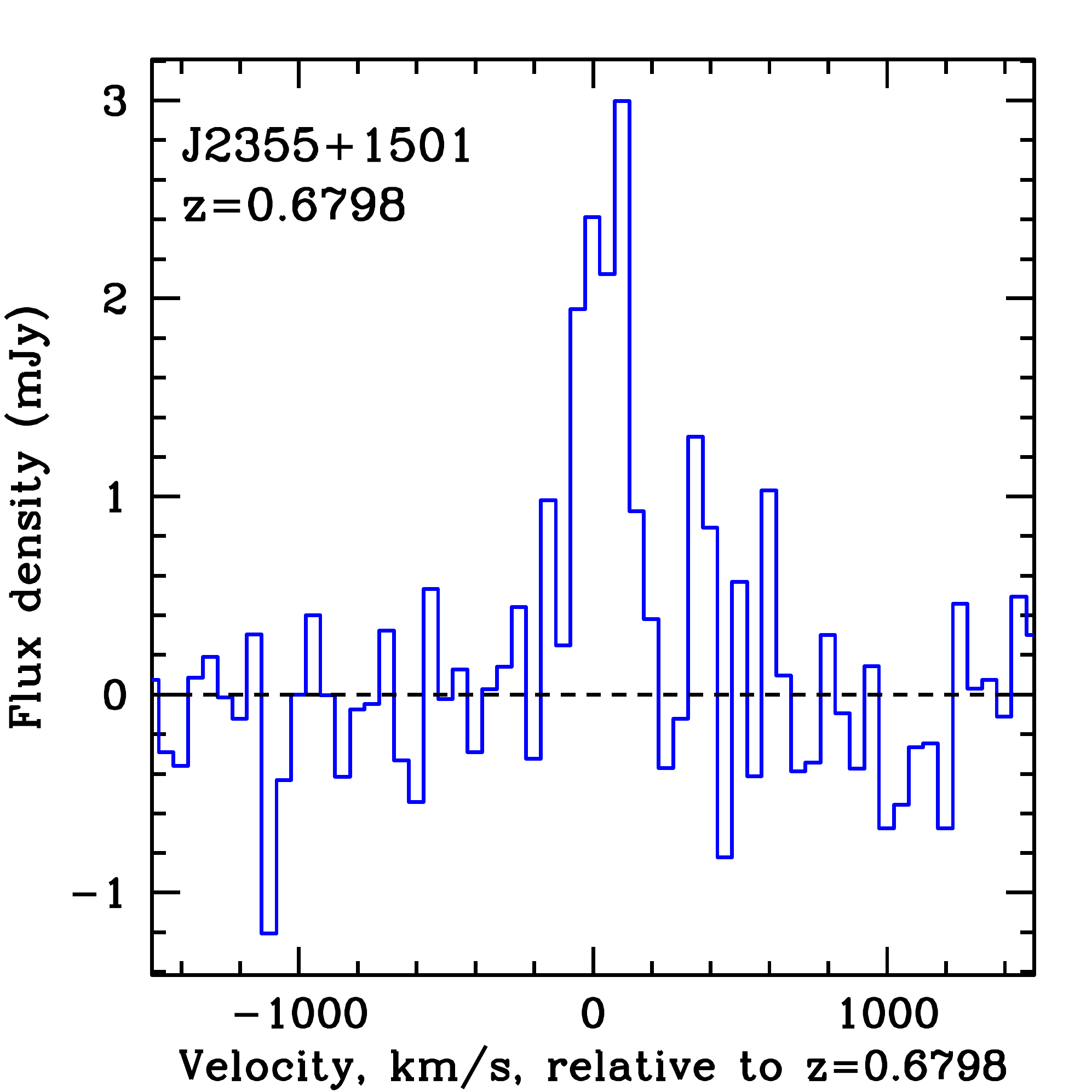}
\includegraphics[width=2.2in]{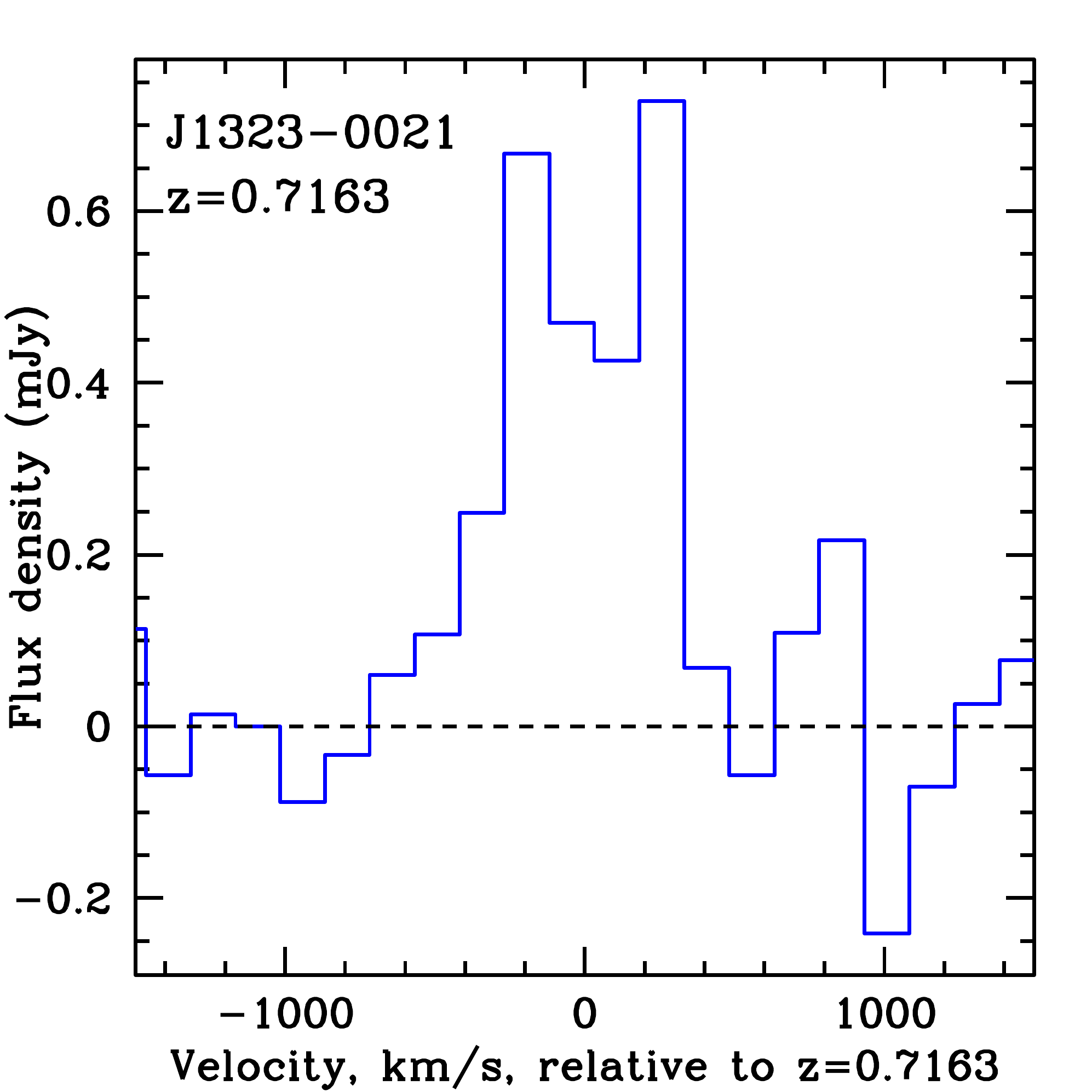}
\includegraphics[width=2.2in]{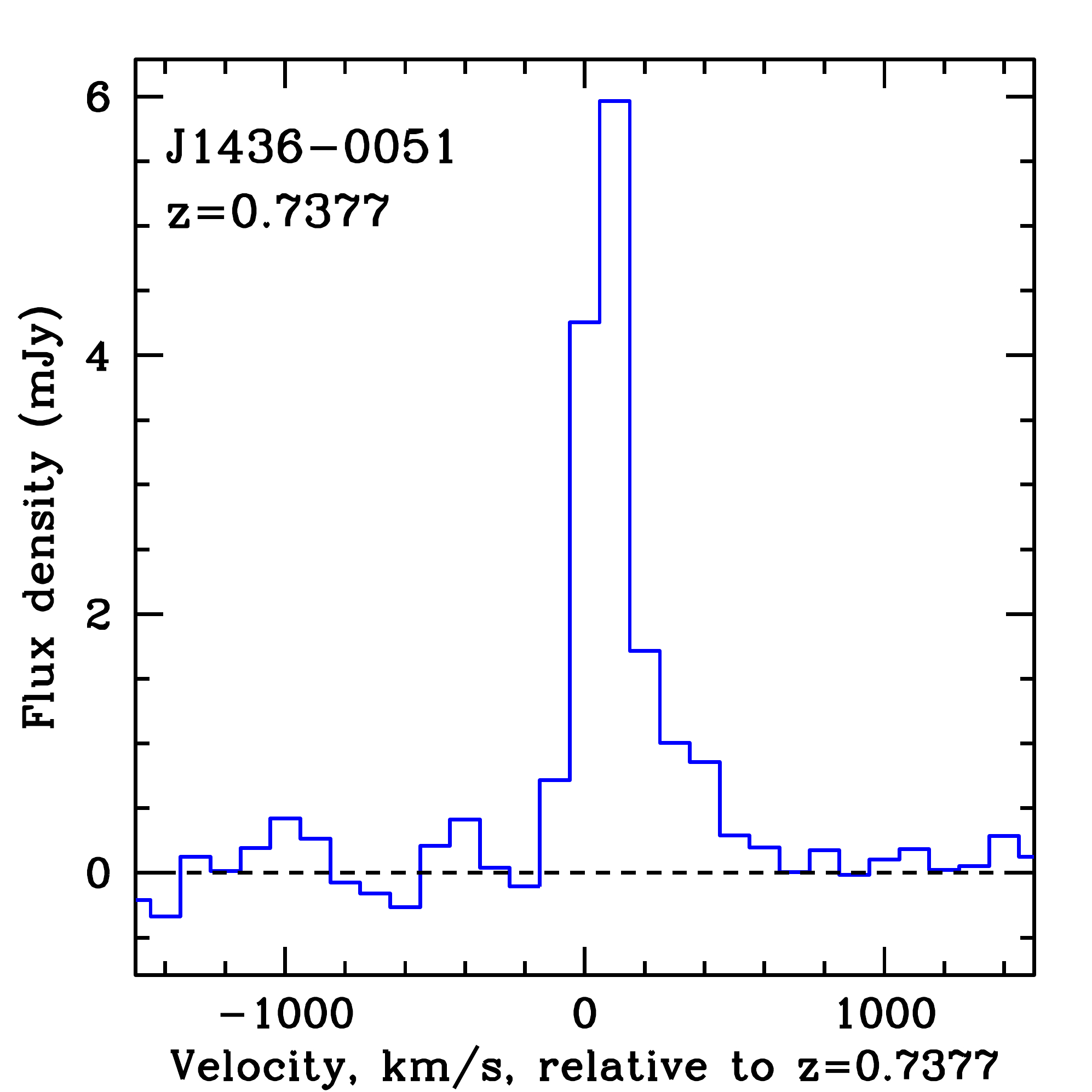}
\caption{CO(2$-$1) line spectra for the five detections. The quasar name and the DLA redshift, 
from low-ionization metal lines, are indicated at the top left corner of each panel. The 
abscissa shows velocity relative to the absorption redshift. The spectra have been binned into 
50~km/s (B0827+243, J2335+1501), 100~km/s (J1436$-$0051) and 150~km/s (B1629+120, J1323-0021) channels.
\label{fig:co}}
\end{figure*}

\begin{figure*}[t!]
\centering
\includegraphics[width=2.2in]{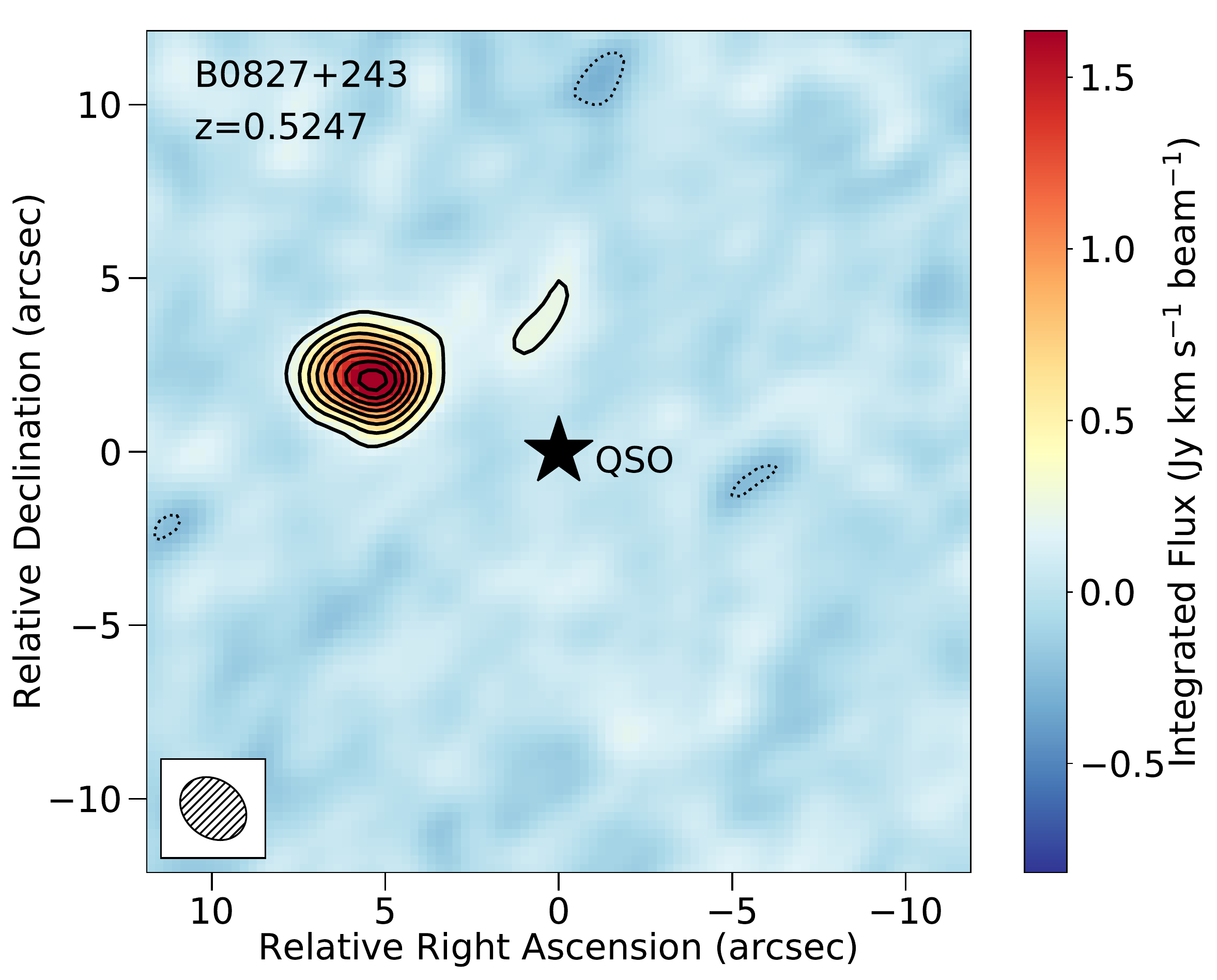}
\includegraphics[width=2.2in]{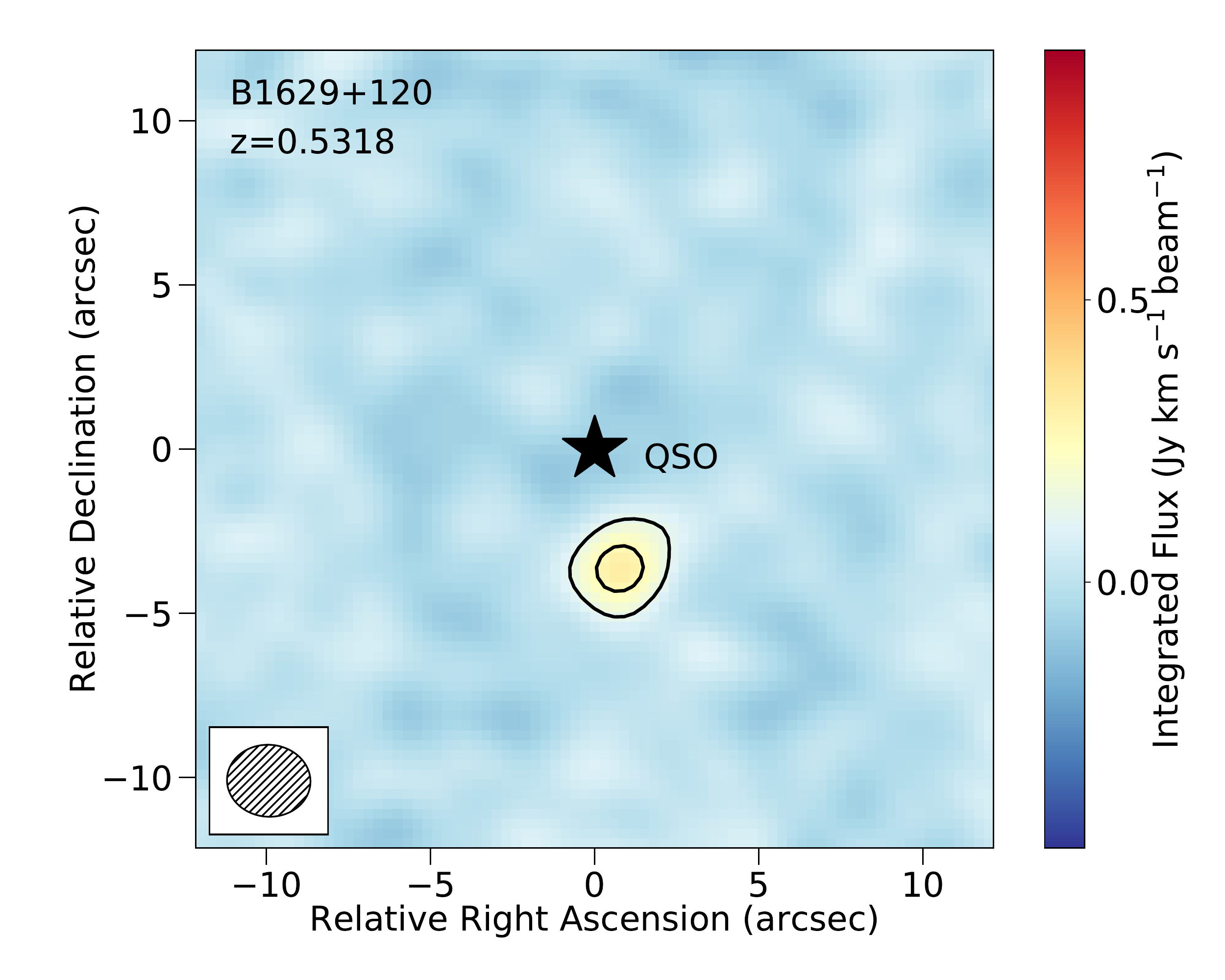}
\includegraphics[width=2.2in]{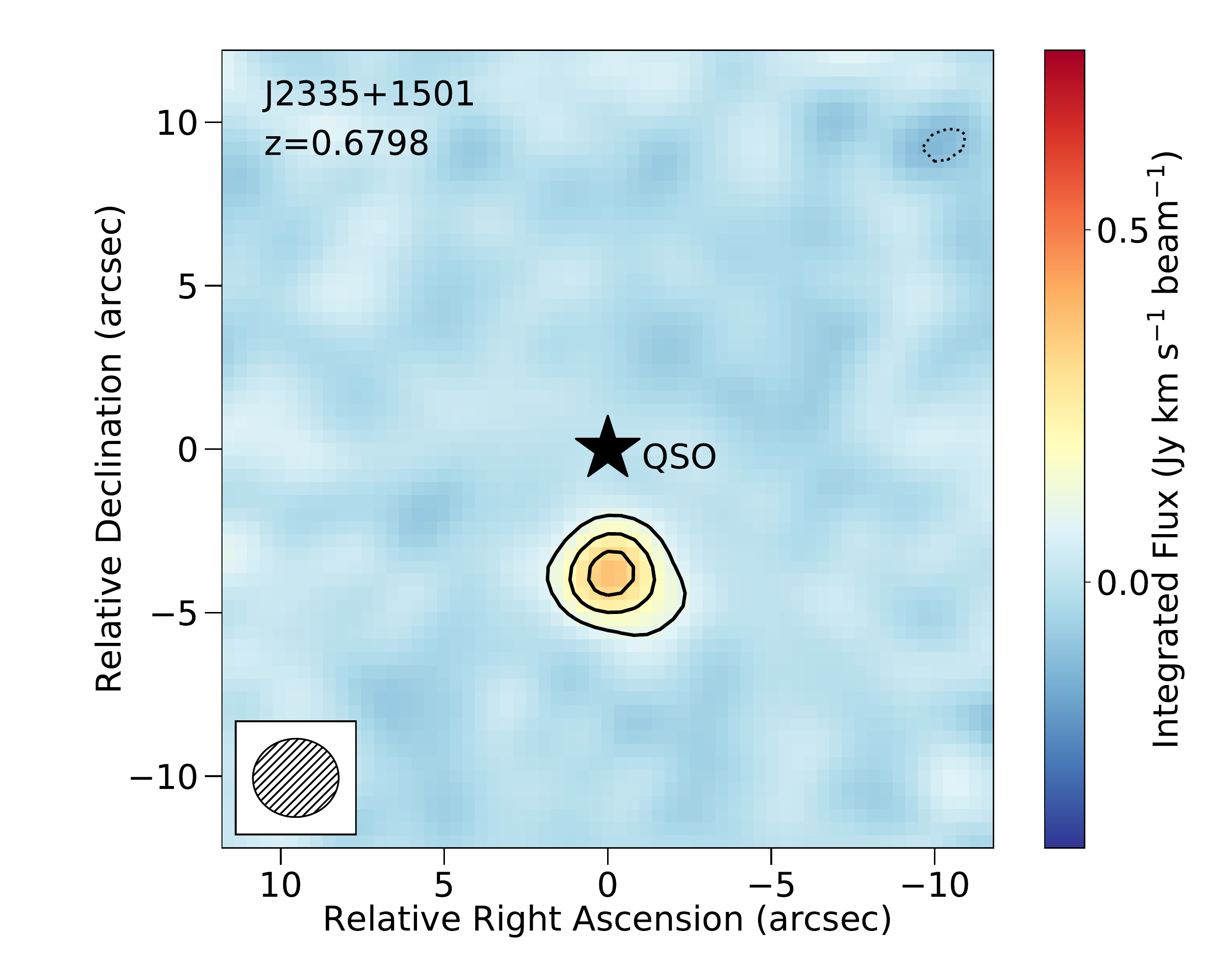}
\includegraphics[width=2.2in]{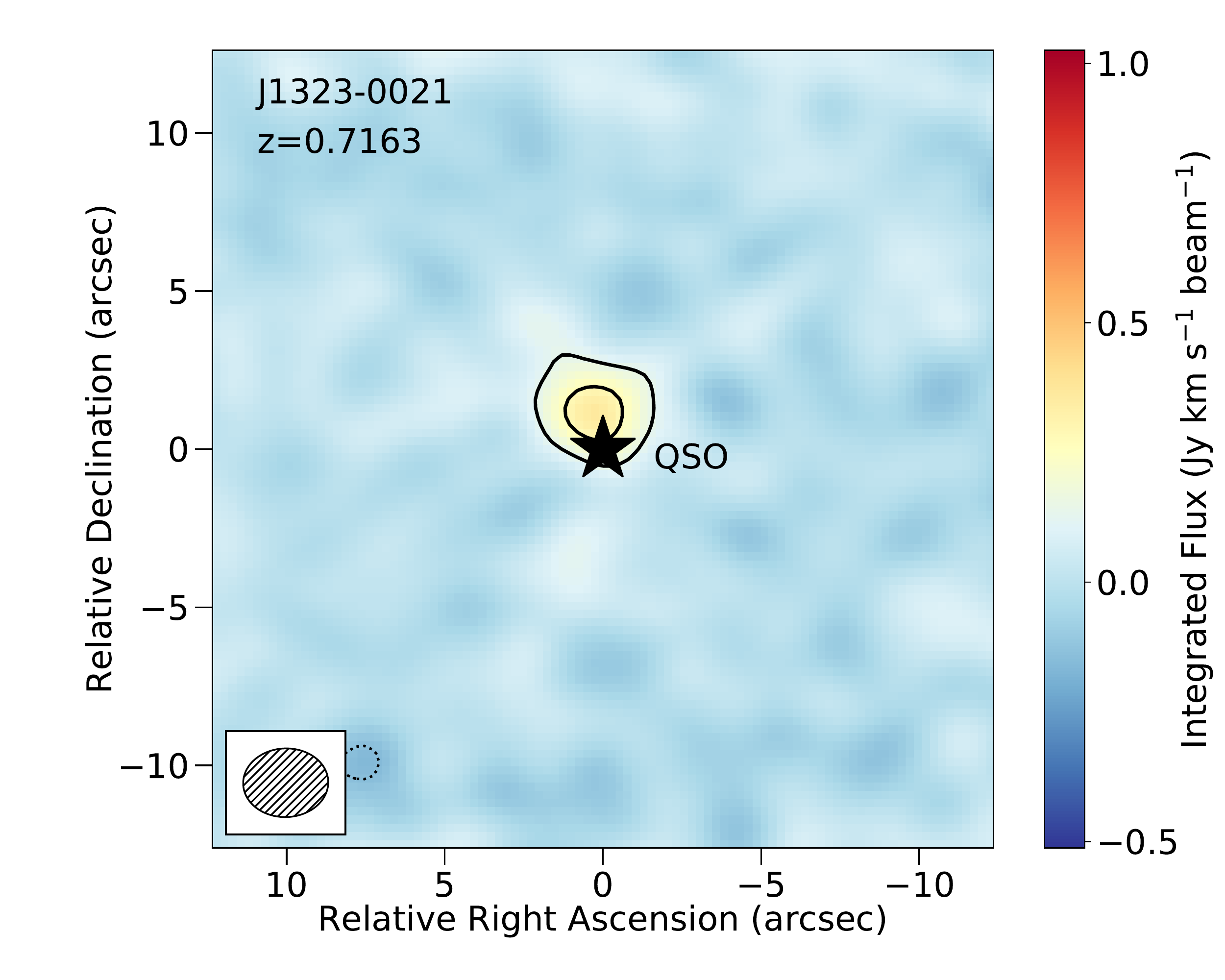}
\includegraphics[width=2.2in]{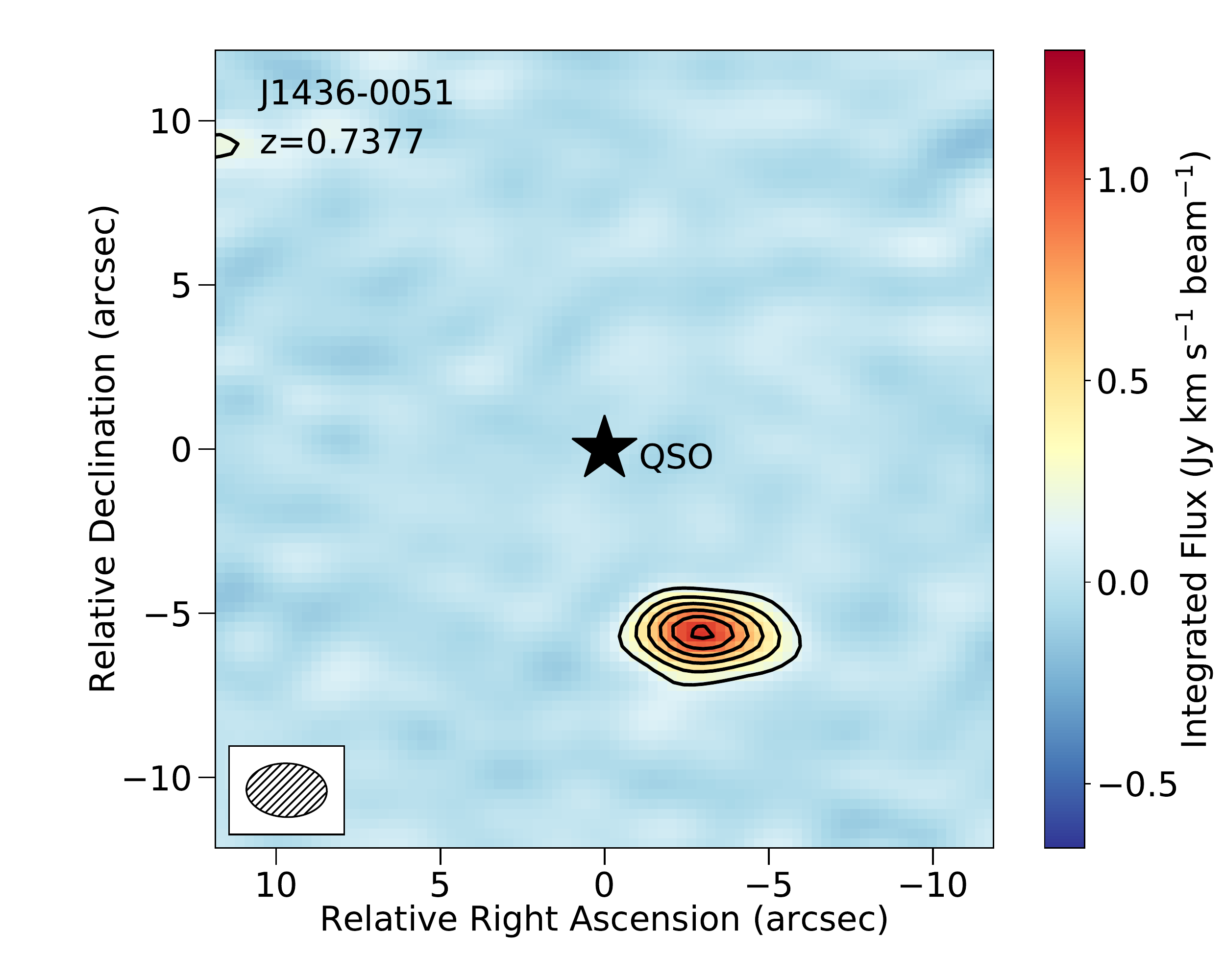}
\caption{Velocity-integrated CO(2$-$1) emission for our five detections; the quasar name and 
the DLA redshift indicated at the top left corner of each panel. The axes co-ordinates are relative 
to the quasar's J2000 co-ordinates. The positive (solid) contours are at $(3,6,9, ...)$~$\times \sigma$, 
and the negative (dashed) contour at $-3\sigma$, where $\sigma$ is the RMS noise on each image. 
The hatched ellipse in each panel shows the ALMA synthesized beam.
	\label{fig:comap}}
\end{figure*}

Our ALMA observations detect strong ($> 6\sigma$ significance) redshifted CO(2$-$1) emission from five of 
our seven targets, spatially coincident with the optically-identified galaxies and at the galaxy redshift. 
Fig.~\ref{fig:co} shows the CO(2$-$1) emission line profiles for the five detections, in order of increasing 
redshift.  The CO lines have full-widths-between-20\%-points, W20,~$\approx 350-650$~km~s$^{-1}$, similar to widths 
observed in nearby galaxies \citep[e.g.][]{saintonge11}. Fig.~\ref{fig:comap} shows the velocity-integrated CO(2$-$1) 
emission maps for the five detections. The spatial extent of the integrated CO(2$-$1) emission was measured 
using the fitter function in the CASA package; the deconvolved sizes are listed in Table~\ref{tab:results}. In four 
of the five cases, the exception being the DLA towards B1629+120, the CO emission is marginally resolved by the 
ALMA synthesized beam, yielding a spatial extent of $\approx 15$~kpc.

\section{Results and Discussion}

\begin{table*}
\centering
\caption{The derived properties of the galaxies associated with our seven intermediate-redshift 
absorbers. 
\label{tab:results}}
\begin{tabular}{|c|c|c|c|c|c|c|c|c|c|}
\hline
DLA galaxy$^a$ & $z_{\rm abs}$$^b$ & M$_{\rm mol}$          & M$_\star$       & SFR    & $f_{\rm Mol}$ & $\tau_{\rm dep}$ & b$^c$ & Deconvolved angular size \\
               &                   & $\times 10^9$~M$_\odot$ & $\times 10^9$~M$_\odot$ & M$_\odot$/yr &               & Gyr              & kpc &  \\
\hline
\hline
DLA~J083052.49+241101.95 & $0.5247$ & $83.1 \pm 4.3$ & $12.3^{+5.1}_{-3.6}$   & $0.7 \pm 0.1$   & $0.87$ & $119$   & $36.3$ & $(1.9 \pm 0.2)'' \times (0.85 \pm 0.37)''$ \\
DLA~J163145.19+115559.40 & $0.5313$ & $6.7 \pm 1.1$  & $9.5^{+2.7}_{-0.1}$    & $1.4 \pm 0.1$   & $0.41$ & $4.8$   & $22.8$ & Unresolved ($< 2.5''$)\\
DLA~J0058+0155           & $0.6125$ & $< 3.3$        & $40.7^{+5.0}_{-4.4}$   & $0.31 \pm 0.02$ & $<0.07$ & $<11$  & $-$    & $-$ \\
DLA~J233544.18+150114.47 & $0.6798$ & $27.3 \pm 2.9$ & $6.8^{+3.1}_{-3.4}$    & $9.5 \pm 0.4$   & $0.80$ & $2.9$   & $26.8$ & $(1.59 \pm 0.57)'' \times (0.79 \pm 0.57)''$ \\
DLA~J132323.82-002153.93 & $0.7163$ & $25.2 \pm 2.5$ & $63.1^{+24.0}_{-17.4}$ & $1.6 \pm 0.6$   & $0.29$ & $15.7$  & $9.6$  & $(1.91 \pm 0.21) \times (0.85 \pm 0.37)''$ \\
DLA~J143644.85-005156.25 & $0.7377$ & $79.3 \pm 2.9$ & $25.7^{+6.7}_{-5.3}$   & $2.2 \pm 0.2$   & $0.76$ & $36.7$  & $45.9$ & $(1.8 \pm 0.2)'' \pm (0.60 \pm 0.21)''$ \\
DLA~J0138-0005           & $0.7821$ & $< 4.9$        & $7.2^{+1.0}_{-1.0}$    & $-$             & $<0.40$  & $-$   & $-$    & $-$ \\
\hline
\end{tabular}
\vskip 0.1in
$^a$~The galaxy name ("DLA" followed by the J2000 co-ordinates for the CO detections, and "DLA" followed by the QSO's J2000 co-ordinates
for the CO non-detections).\\
$^b$~The absorber redshift, from low-ionization metal lines.\\
$^c$~The impact parameter, in kpc,  at the absorber redshift, to the quasar line of sight.
\vskip 0.1in
\end{table*}

The molecular gas mass of a galaxy can be inferred from its CO(2$-$1) line luminosity 
if one knows the CO-to-H$_2$ conversion factor ($\alpha_{\rm CO}$) and the nature of the 
excitation of the J$=2$ level \citep[e.g.][]{bolatto13,carilli13}. For high-metallicity galaxies,
like spiral disks and starbursts, the CO-to-H$_2$ conversion factor is typically low, 
$\alpha_{\rm CO} \approx 4.36$~M$_\odot$~(K~km/s~pc$^{2}$)$^{-1}$ (spirals) and 
$\approx 1$~M$_\odot$~(K~km/s~pc$^{2}$)$^{-1}$ (starbursts) \citep{bolatto13}. To allow for 
a direct comparison with emission-selected galaxies in the local Universe \citep[the 
{\it COLD~GASS} sample; e.g.][]{saintonge11} 
and at high redshifts \citep[the {\it PHIBSS} sample; e.g.][]{tacconi13}, we 
will throughout use $\alpha_{\rm CO} \approx 4.36$~M$_\odot$~(K~km/s~pc$^{2}$)$^{-1}$.
For the line excitation, we assume sub-thermal excitation of the CO J$=2$ level, as 
is typical in normal spiral galaxies in the local Universe and colour-selected galaxies 
at high redshifts. In the Milky Way and nearby spirals, the J$=2-1$ line is $\approx 2.5$ 
times stronger than the J$=1-0$ line \citep[e.g.][]{fixsen99,weiss05}; we will assume this value 
for the ratio of the line flux densities in order to infer the CO(1$-$0) line luminosity, 
$L'_{\rm CO(1-0)}$. The inferred molecular gas masses for the galaxies associated with the five 
absorbers with CO(2$-$1) line detections lie in the range $(0.6-8) \times 10^{10} \times 
(\alpha_{\rm CO}/4.36) \times (0.63/r_{21})$~M$_\odot$.

The stellar mass (M$_\star$) of the galaxies associated with six of our seven targets was 
estimated by fixing the redshift to the spectroscopic value, and matching spectral energy 
distribution (SED) templates to multi-band photometry using {\tt HyperZ} \citep{bolzonella00} 
and {\tt LePhare} \citep{arnouts99,ilbert06}. We use standard BC03 single stellar population 
spectral templates \citep{bruzual03}, assuming a Chabrier initial mass function. M$_\star$ is 
determined by the parameter-set that minimises the $\chi^2$-statistic across a user pre-defined 
grid of models. The grid encompasses ages and e-folding time-scales in the ranges 0.01-13.5 Gyrs 
and 0.1-30 Gyrs respectively, and accounts for intrinsic extinction assuming a standard LMC-Fitzpatrick 
law. To parametrize the amount of extinction, {\tt HyperZ} accepts reddening as a free parameter,
which we vary in the range $A_V = 0-1$ mag, whereas {\tt LePhare} accepts a set of user-defined 
$E_{B-V}$ values, which we select to be in the range $0-1$. 
We have verified that {\tt HyperZ} and {\tt LePhare} produce consistent stellar mass estimates, 
within the uncertainties. We obtain stellar masses in the range $M_\ast \approx (0.7-6.5) \times 10^{10} \: {\rm M}_\odot$, 
listed in Table~\ref{tab:results}. Note that the stellar mass of the galaxy associated with the $z=0.7821$ 
sub-DLA towards J0138-0005 is based on a conversion from its K-band magnitude to the stellar mass 
\citep{cappellari13}.

Finally, the star formation rates (SFRs) of the DLA galaxies were inferred from either the [O{\sc ii}]$\lambda$3727\AA\ 
or H$\beta$$\lambda$4382\AA\ lines \citep[from the literature; e.g.][]{chen05,christensen14,straka16,moller18,rhodin18}, 
or (in the case of the DLA towards B1629+120), from the $U$-band continuum \citep{rao03}, using standard SFR calibrations 
\citep{kennicutt12}. We do not have an estimate of the SFR for the galaxy associated with the $z= 0.7821$ sub-DLA towards 
J0138-0005. While these tracers may under-estimate the SFR for dusty galaxies, they allow a fair comparison
to the emission-selected samples, where the SFR estimates are based on the [O{\sc ii}]$\lambda$3727\AA\ or 
H$\alpha$~$\lambda$6563\AA\ emission line luminosity. The SFRs are relatively low, $\approx (0.3-9.5) \: 
{\rm M}_\odot$~yr$^{-1}$, again tabulated in Table~\ref{tab:results}.

The two absorbers with non-detections of CO(2$-$1) emission, at $z=0.6125$ towards J0058+0155 and $z=0.7821$ towards 
J0138-0005 do not stand out in their absorption properties from the rest of the sample. We note, in passing,
that the galaxy counterpart of the $z = 0.7821$ sub-DLA does not have a spectroscopic redshift, 
while the SFR of the $z = 0.6125$ sub-DLA is the lowest in our sample ($0.31 {\rm M}_\odot$~yr$^{-1}$), 
despite its relatively high stellar mass (M$_\star \approx 4 \times 10^{10}$~M$_\odot$).

Our estimates of the molecular gas mass in the galaxies associated with our seven absorbers allow us
to compare the gas properties of high-metallicity, absorption-selected galaxies to those of 
emission-selected, star-forming galaxies, to test whether the same galaxy populations are being 
selected by the different methods. Unfortunately, there are few emission-selected galaxies with CO studies 
at intermediate redshifts, $z \approx 0.7$. For the comparison, we hence used two large emission-selected 
galaxy samples, the {\it COLD GASS} sample at low redshifts ($z \lesssim 0.05$), with 212 CO(1$-$0) detections
in galaxies with stellar mass~${\rm M}_\ast \geq 10^{10} \: {\rm M}_\odot$ \citep{saintonge11},
and the {\it PHIBSS} sample at high redshifts ($z \approx 1.3,2.2$) with 49 CO(3$-$2) detections 
in galaxies with SFR~$\geq 30 \: {\rm M}_\odot$~yr$^{-1}$ and stellar 
mass~$\rm M_\ast \geq 2.5 \times 10^{10} M_\odot $ \citep{tacconi13}. We assume 
$\alpha_{\rm CO} = 4.36 \: {\rm M}_\odot$~(K~km~s$^{-1}$~pc$^2$)$^{-1}$ for all galaxies.

\begin{figure*}[t!]
\centering
\includegraphics[width=3.0in]{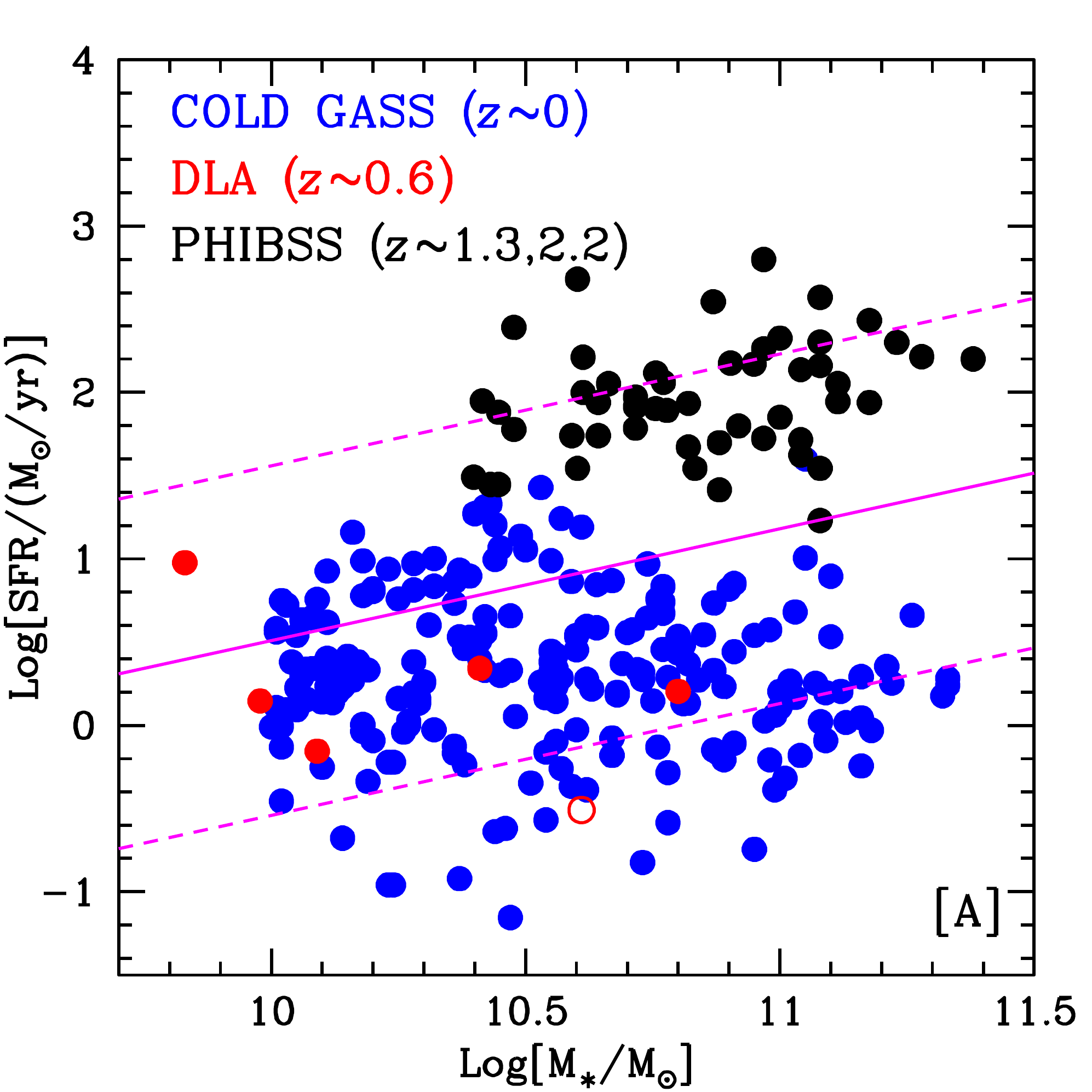}
\includegraphics[width=3.0in]{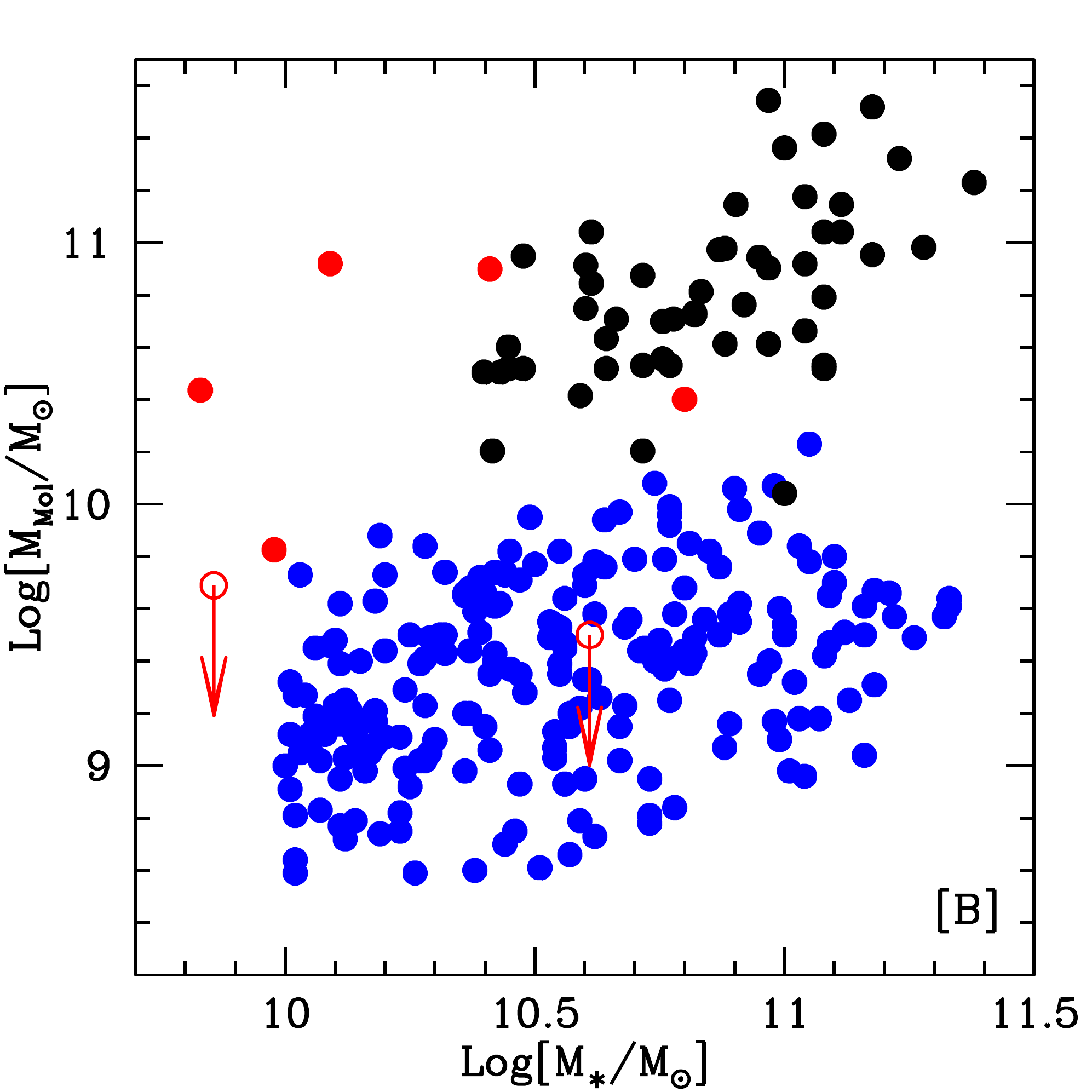}
\includegraphics[width=3.0in]{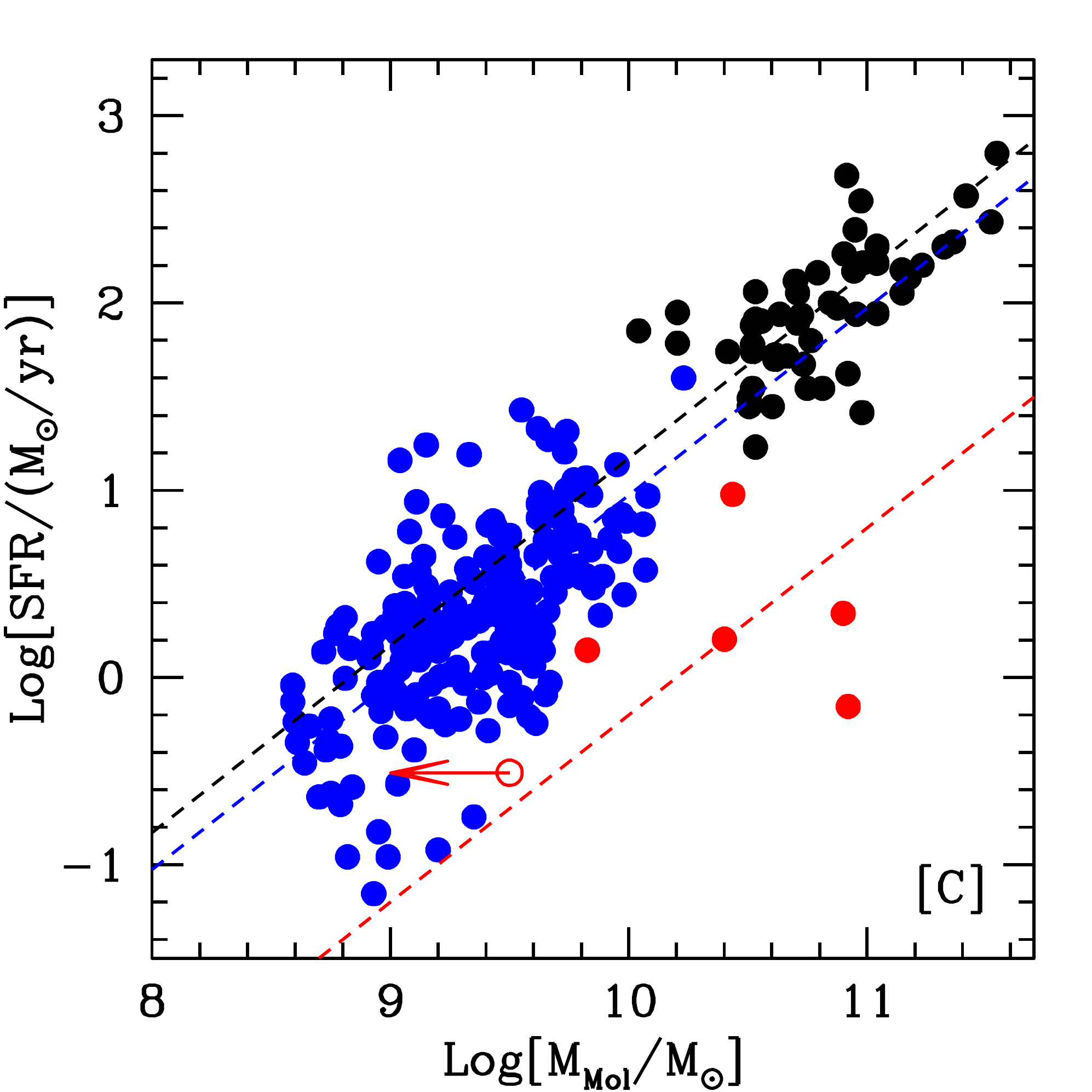}
\includegraphics[width=3.0in]{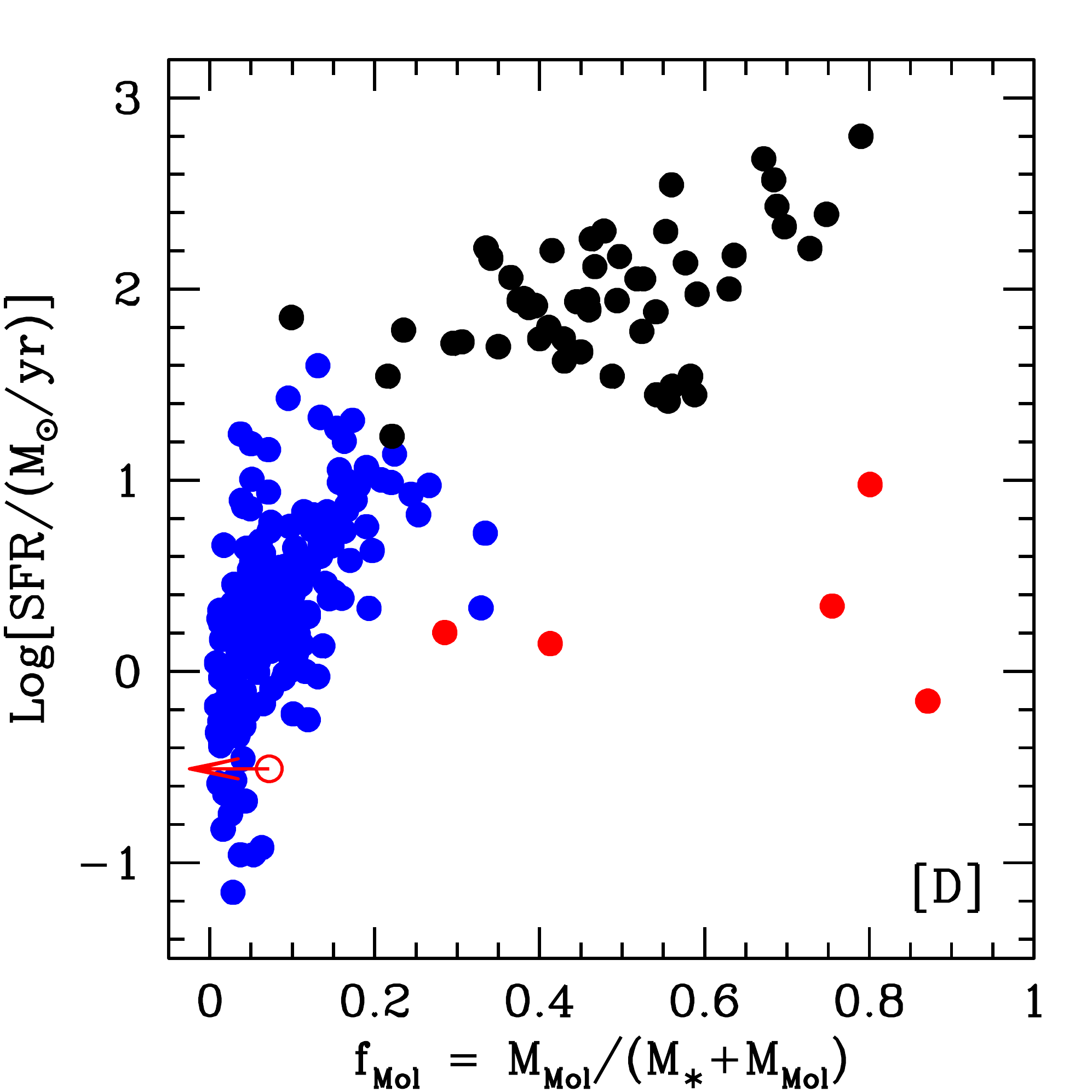}
\caption{Comparisons between various stellar and gas properties of the galaxies associated with intermediate-redshift 
absorbers (shown in red) with those of two large emission-selected galaxy samples, the {\it COLD~GASS} sample at 
$z \approx 0$ (in blue; \citep{saintonge11}) and the {\it PHIBSS} sample at $z \approx 1.3,2.2$ (in black; \citep{tacconi13}). 
[A]~The SFR plotted against the stellar mass: the solid magenta line shows the main-sequence relation of 
star-forming galaxies at $z \approx 0.2-0.7$, with the dashed magenta lines showing the $\pm 3\sigma$ scatter in 
the relation \citep{noeske07}. [B]~The molecular gas mass plotted versus the stellar mass. 
[C]~The SFR plotted versus the molecular gas mass; the three dashed lines in the figure indicate the median gas depletion 
times for the three samples. [D]~The SFR plotted versus the molecular gas fraction. See text for discussion.
\label{fig:compare}}
\end{figure*}

The four panels of Fig.~\ref{fig:compare} show comparisons between different properties of 
the DLA (red), {\it COLD GASS} (blue) and {\it PHIBSS} (black) samples. The top-left panel
shows the optical properties, with the SFR plotted against the stellar mass: the DLA galaxies 
appear consistent with being drawn from the population of star-forming galaxies at $z \approx 0.2-0.7$ 
\citep{noeske07} and also with the {\it COLD GASS} population. The top-right panel plots the molecular gas mass 
versus the stellar mass: four of the seven DLA galaxies have substantially higher molecular gas masses 
than the galaxies of the low-$z$ {\it COLD GASS} sample, at the same stellar masses. The bottom-left panel 
shows the SFR plotted against the molecular gas mass, with the dashed lines indicating the median 
molecular gas depletion times ($\rm \tau_{\rm depl} =~{\rm M}_{Mol}/{\rm SFR}$) for the three samples.
The {\it PHIBSS} and {\it COLD GASS} galaxies show a clear correlation between SFR and molecular gas mass 
\citep{saintonge11b,tacconi13}, while the DLA galaxies are offset from this correlation, 
with far lower SFRs at the same molecular gas masses when compared to the {\it PHIBSS} galaxies, 
and higher molecular gas masses at the same SFRs, relative to the {\it COLD GASS} galaxies. 
The fact that the DLA galaxies lie below the correlation between SFR and molecular gas mass suggests
that star formation activity is being quenched in these systems. Consistent with this picture, the 
median gas depletion time for the DLA galaxies is $\rm \tau_{\rm depl,DLA} \approx 10$~Gyr, a factor 
of $\approx 10$ larger than the median gas depletion times in the {\it PHIBSS} and {\it COLD GASS} samples 
($\approx 0.7$~Gyr and $\approx 1$~Gyr, respectively). Finally, the bottom right panel shows the 
SFR plotted against the molecular gas fraction, $f_{\rm Mol} = \: {\rm M_{Mol}}/({\rm M}_\ast + {\rm M_{Mol}})$: 
for the {\it COLD GASS} and {\it PHIBSS} galaxies, the SFR increases with increasing gas fraction, 
while the DLA galaxies lie well below this correlation, with substantially lower SFRs at the same 
molecular gas fraction, or far higher molecular gas fractions at the same SFR.

From Fig.~\ref{fig:compare}, the molecular gas properties of galaxies associated with high-metallicity 
DLAs at $z \approx 0.7$ appear distinct from those of star-forming galaxies in the local Universe 
and at high redshift, despite their similar optical properties. The molecular gas masses, gas depletion 
times, and molecular gas fractions are substantially larger, at the same SFRs, for the absorption-selected 
galaxies than for the emission-selected galaxies. Our primary selection criterion for the absorption sample, i.e., 
a high metallicity, allows us to select galaxies similar in stellar mass to those of the {\it PHIBBS} 
and {\it COLD GASS} samples due to the known correlation between metallicity and stellar mass in both 
star-forming galaxies and DLAs \citep{tremonti04,moller13}. Note that the stellar masses of the 
DLA galaxies are similar to those of the emission-selected galaxies, with all the DLA galaxies lying 
within the range of the star-forming main sequence at $z \approx 0.2-0.7$ \citep{noeske07}.

We have assumed the same CO-to-H$_2$ conversion factor, $\alpha_{\rm CO} = 4.36 ~{\rm M}_\odot$
~(K~km~s$^{-1}$~pc$^2$)$^{-1}$ for all galaxies. A low CO-to-H$_2$ conversion factor in DLA galaxies, 
$\alpha_{\rm CO} \approx 1 $, would lower their inferred molecular gas masses, gas fractions, and depletion 
times by a factor of $\approx 5$, reducing some of the tension between the DLAs and the emission-selected samples. 
However, such low values of $\alpha_{\rm CO}$ are only seen in starburst galaxies like ultra-luminous 
infrared galaxies (ULIRGs) \citep{bolatto13}. A low $\alpha_{\rm CO}$ value in DLAs would imply that 
high-metallicity, absorption-selected galaxies at intermediate redshifts are ULIRGs, rather than 
normal star-forming galaxies. This would be surprising as nothing in the optical images of the DLA 
galaxies suggests that these are ULIRGs \citep[e.g.][]{rao03,chen05}, and it is not clear why the absorption 
selection should predominantly yield molecule-rich, starburst galaxies. We hence do not consider this 
possibility to be likely. 

The second possibility is that our selection based on high-metallicity absorption indeed 
picks out typical star-forming galaxies at intermediate redshifts. The normal SFR and stellar 
mass properties, but distinct molecular gas properties, may then indicate a transition in the 
nature of star formation, evolving from the large CO reservoirs and high SFRs at the epoch 
of peak star formation ($z \approx 1-3$) to lower SFRs but similarly large 
molecular masses at $z \approx 0.7$. This might arise if most of the dense molecular gas in 
these galaxies has been consumed in the process of star formation, and the bulk of the 
remaining molecular gas is at a low density, insufficient to trigger further star formation 
(e.g.,  a ``post-starburst'' scenario). Such extended diffuse molecular disks would have a 
large atomic gas cross-section, making them relatively easy to detect in damped Lyman-$\alpha$ 
absorption. A recent ALMA search for CO emission from two massive post-starburst galaxies 
at $z \approx 0.7$ with quenched star formation yielded large molecular gas masses, 
similar to our values \citep{suess17}. 

While our sample is yet small, our results suggest that high-metallicity galaxies at intermediate 
redshifts, $z \approx 0.7$, selected via their absorption signatures, have very different molecular 
gas properties (higher masses, gas fractions, and depletion times) from emission-selected star-forming 
galaxies both at the peak epoch of star formation in the Universe and at low redshifts, $z < 0.05$. 
The large inferred molecular gas reservoirs but low levels of star formation suggest that either 
high-metallicity, absorption-selected galaxies are dusty, starburst systems, or there is a transition 
in the nature of star formation in galaxies at intermediate redshifts, $z \approx 0.7$.

\acknowledgements
NK acknowledges support from the Department of Science and Technology via a Swarnajayanti Fellowship 
(DST/SJF/PSA-01/2012-13). JXP acknowledges support from NSF AST-1412981. LC and NHPR are supported by 
DFF - 4090-00079. Support for this work was provided by the NSF through award SOSPA2-002 
from the NRAO. ALMA is a partnership of ESO (representing its member states), NSF (USA) and 
NINS (Japan), together with NRC (Canada), NSC and ASIAA (Taiwan), and KASI (Republic of Korea), 
in cooperation with the Republic of Chile. The Joint ALMA Observatory is operated by ESO, AUI/NRAO 
and NAOJ. The data reported in this paper are available though the ALMA archive 
(https://almascience.nrao.edu/alma-data/archive) with project codes: ADS/JAO.ALMA \#2013.1.01178.S and 
\#2015.1.01034.S.


\end{document}